\documentclass[%
 reprint,
superscriptaddress,
 amsmath,amssymb,
aps,
]{revtex4-2}
\usepackage{textcomp} 
\usepackage[normalem]{ulem}
\usepackage{CJKutf8}
\usepackage{chemfig}
\usepackage{lineno,hyperref}
\usepackage[squaren]{SIunits}
\usepackage{color}
\usepackage{amsmath}
\usepackage{caption,subcaption}
\usepackage{setspace}
\usepackage{multirow}
\usepackage{threeparttable}
\usepackage{booktabs}
\usepackage{graphicx}% Include figure files
\usepackage{dcolumn}% Align table columns on decimal point
\usepackage{bm}% bold math

\definecolor{americanrose}{rgb}{1.0, 0.01, 0.24}
\definecolor{coralpink}{rgb}{0.97, 0.51, 0.47}
\definecolor{ao(english)}{rgb}{0.0, 0.5, 0.0}
\definecolor{darkpastelgreen}{rgb}{0.01, 0.75, 0.24}
\definecolor{cyan(process)}{rgb}{0.0, 0.72, 0.92}

\newcommand\abctol{S1}
\newcommand\hytem{S2}

\newcommand\chemNN{--$\stackrel{|}{\text{N}}$--CO--$\stackrel{|}{\text{N}}$--}

\newcommand\chemNC{--$\stackrel{|}{\text{N}}$--CO--$\underset{|}{\stackrel{|}{\text{C}}}$--}

\begin{document}

\preprint{APS/123-QED}

\title{Temperature and Tautomeric Effects in High-Resolution Oxygen 1s X-ray Photoelectron Spectroscopy of Purines and Pyrimidines}

%of heterocyclic biochemicals at the oxygen K-edge}
%\title{First-Principles Simulations of Vibrationally-Resolved O1s X-ray Photoelectron Spectra of Purines  and  Pyrimidines: Insights into the Carbonyl Bonding and Temperature/Tautomeric Effects}

%\title{Franck-Condon simulations of vibrationally-resolved X-ray photoelectron spectra of heterocyclic biochemicals at the oxygen K-edge}
%\title{Full core hole calculations of vibrationally-resolved X-ray photoelectron spectra of  molecules at the N K-edge}% Force line breaks with \\
% as influenced by consecutive CH$\leftrightarrow$N replacement

\author{Minrui Wei}
 \affiliation{MIIT Key Laboratory of Semiconductor Microstructure and Quantum Sensing, Department of Applied Physics, School of Physics, Nanjing University of Science and Technology, 210094 Nanjing, China}%Lines break automatically or can be forced with 
 
\author{Junxiang Zuo}
\email{jxzuo@njust.edu.cn}
\affiliation{MIIT Key Laboratory of Semiconductor Microstructure and Quantum Sensing, Department of Applied Physics, School of Physics, Nanjing University of Science and Technology, 210094 Nanjing, China}
 
%\author{Lu Zhang}
% \affiliation{MIIT Key Laboratory of Semiconductor Microstructure and Quantum Sensing, Department of Applied Physics, School of Physics, Nanjing University of Science and Technology, 210094 Nanjing, China} 

\author{Guangjun Tian}%
% \email{tian@ysu.edu.cn}
\affiliation{Key Laboratory for Microstructural Material Physics of Hebei Province, School of Science, Yanshan University, Qinhuangdao 066004, China}%
 
\author{Weijie Hua}%
 \email{wjhua@njust.edu.cn}
 \affiliation{MIIT Key Laboratory of Semiconductor Microstructure and Quantum Sensing, Department of Applied Physics, School of Physics, Nanjing University of Science and Technology, 210094 Nanjing, China}%Lines break automatically or can be forced with \\

\date{\today}% It is always \today, today,
%\begin{CJK*}{UTF8}{gbsn}
\begin{abstract}
Purines and pyrimidines, crucial building blocks in biological systems, have attracted significant interest across molecular physics, biochemistry, pharmacology, and chemistry.  Extensive spectroscopies have been employed for characterization, while the temperature and potential tautomeric effects can complicate the interpretation of underlying physics and chemistry. Here, we conducted first-principles simulations to analyze the vibrationally-resolved O1s X-ray photoelectron spectra (XPS) of six common biomolecules at different temperatures, comprising three purine  (xanthine, caffeine, and hypoxanthine) and three pyrimidine (thymine, 5F-uracil, and uracil) derivatives, and the tautomeric effect of hypoxanthine at varying temperatures.  Using both time-independent (TI) and time-dependent (TD) methods under the Franck-Condon approximation, we obtained theoretical spectra that exhibited excellent agreement with experiments. Our analysis of these systems, all featuring carbonyl oxygens, unveiled distinctive characteristics of oxygen in the local structure of {\chemNN} (O2) compared to that within a {\chemNC} structure (O1), showcasing higher O1s binding energy and total vibrational reorganization energy. We observed small differences ($\Delta$ZPE) in the zero-point vibration energies between the core-ionized and ground states, indicating a weak Duschinsky rotation effect. Through structural analysis, we consistently found that ionization of the O1s electron resulted in elongation of the O$^*$=C bond length (by 0.08--0.09 \AA) for this specific category of molecules. The TI method facilitated the assignment of experimental spectra to different atoms or tautomers, where the atom-specific vibronic profiles of all six molecules exhibited similarity, with the 0-2 transitions dominating. Meanwhile, the TD method enabled a more comprehensive exploration of the temperature effect, and the tautomeric effect (specifically for hypoxanthine) by incorporating the Boltzmann population ratios of tautomers. Notably, we observed significant temperature dependence in the vibronic features present in these spectra.

  \end{abstract}
  
\maketitle

%---------------------------------------------------------------------------------
\section{Introduction} 
Small heterocyclic biochemical compounds, including derivatives of purine and pyrimidine, play a crucial role in biological systems and have attracted significant attention in the fields of molecular physics, biochemistry, pharmacology, and chemistry.  For example, essential nucleic acid bases such as uracil and thymine are fundamental components involved in the transmission of genetic information transmission,\cite{lazcano_evolutionary_1988} biochemical reaction mechanisms,\cite{parikh_lessons_2000, beukers_50_2008} pharmaceutical design,\cite{koch_thymine_2017, ramesh_therapeutic_2020} and disease research.\cite{noauthor_recent_2020, ramesh_therapeutic_2020} These compounds provide a vital foundation for understanding life processes and developing innovative therapeutic approaches, as evidenced by their applications in oncological therapeutics, with 5F-uracil being a well-known agent.\cite{noauthor_recent_2020} Xanthine, on the other hand, is a fundamental component in various natural substances and is found in numerous pharmacologically active compounds.\cite{kapri_recent_2022} Its derivatives exhibit diverse physiological and pharmacological activities. Notably, caffeine, derived from xanthine, acts as a central nervous system stimulant.\cite{ferre_update_2008}  Hypoxanthine sensors have significant contributions to various fields including food safety, medical diagnosis, and biomedical research.\cite{garg_review_2022} Therefore, it is essential to model the structures of these small heterocyclic biochemical compounds to anticipate their behavior within larger molecules and understand their implications in various contexts.

X-ray photoelectron spectroscopy (XPS) has emerged as a powerful tool for investigating chemical bonds and electronic structures of molecules.\cite{delesma_chemical_2018, andrade_x-ray_1985, ketenoglu_general_2022, sokolowski1957magnetic, nordling_precision_1957, siegbahn1969esca} The measurement of core electron binding energy (BE) in XPS serves as a localized probe, providing valuable insights into the chemical environment of atoms within a molecule. This allows for the establishment of structure-property relationships among different compounds. The development of XPS technology can be attributed to the pioneering work of Kai Siegbahn,\cite{siegbahn1982electron, siegbahn1967atomic, siegbahn1974esca} and there have been consistent improvements in detection resolutions.  High-resolution vibrationally-resolved XPS spectra offer more detailed information beyond just binding energies.\cite{carravetta_x-ray_2022, hergenhahn_vibrational_2004, svensson_soft_2005}  They provide insights into the vibronic fine structure, which reflects potential energy surface (PES) information related to the core-electron ionized and ground states near the Franck-Condon (FC) region. Additionally, these spectra capture the dynamics of electrons and nuclei induced by the core ionization, thus offering a deeper understanding of molecular behavior. 

Excitation of polyatomic molecules is often accompanied by multiple vibrational modes, resulting in complex spectra.  Theoretical simulations play an important role in interpreting the fine structures observed in experiments. In this regard, approaches that combine full core hole (FCH) density functional theory (DFT) with the Duschinsky rotation (DR) method, under the harmonic oscillator approximation, have proven effective for polyatomic molecules.\cite{hua_theoretical_2020, cheng_vibrationally-resolved_2022, wei_vibronic_2022, wei_vibronic_2023, couto_breaking_2021}  The time-independent (TI) framework, particularly applicable when the temperature is negligible, offers a computationally efficient and highly precise approach. It allows for the clear interpretation of individual transitions and has shown excellent agreement with experimental XPS spectra for various molecules\cite{hua_theoretical_2020, cheng_vibrationally-resolved_2022, wei_vibronic_2022, wei_vibronic_2023} and ions.\cite{du_theoretical_2022} However, when temperature effects become significant, addressing them within the TI framework introduces additional computational complexity.\cite{santoro_effective_2007} On the other hand, the time-dependent (TD) framework provides an efficient solution by avoiding explicit calculation of multidimensional Franck-Condon factors (FCFs). This makes TD methods highly suitable for obtaining fully converged spectra, especially when considering specific temperatures.\cite{berger_calculation_1998, tian2013electron}

In recent years, significant progress has been made in the investigation of inner shells of small gaseous biological molecules.\cite{plekan_theoretical_2008, plekan_x-ray_2012, feyer_tautomerism_2009, castrovilli_experimental_2018, bolognesi_pyrimidine_2010} Researchers have utilized gas-phase exploration to analyze the spectra of these molecules in a clean environment, free from solvation effects. This approach allows for more precise spectral analysis and provides valuable data for comparison with theoretical simulations. 

In this study, we conducted an in-depth analysis of a specific subset of six molecules: xanthine, caffeine, thymine, 5F-uracil, uracil, and hypoxanthine. Our focus was on exploring the O1s edge in these biomolecules. The presence of oxygen in their side chains plays a pivotal role in various biological processes, such as base pairing,\cite{sivakova_nucleobases_2005} structural stability,\cite{rees_base_2018, kryachko_theoretical_2001} drug activity,\cite{ramesh_therapeutic_2020, faudone_medicinal_2021} and chemical reactions.\cite{mehmood_natural_2019, boncel_uracil_2008, kwasniewska-sip_chemical_2021} Understanding the vibrational properties of these molecules through spectroscopic simulations enables us to gain comprehensive insights into X-ray physics within specific chemical environments. This knowledge contributes to a deeper understanding of the behavior of these molecules and their interactions at the atomic level.

In data interpretation and analysis for various systems, traditional methods that are temperature-independent and performed within the TI framework are typically relied upon. These methods provide a clear understanding of individual spectra transitions. However, it is important to recognize that temperature plays a significant role in determining the structural stability of biomolecules. Temperature can have a profound impact on complex systems, affecting the relative distribution of different tautomers through thermally induced interconversion. This thermal effect is directly reflected in the spectral characteristics of these biomolecules.\cite{li_gas-phase_2014, santoro_effective_2008, plekan_x-ray_2012} XPS enables researchers to gain valuable information about the chemical composition and electronic structure of tautomeric species and identify different tautomeric forms based on their characteristic binding energies and peak intensities. Moreover, changes in the XPS spectra with temperature variations can offer further insights into the temperature-dependent behavior of tautomeric equilibria. By monitoring the intensity or position of specific peaks in the XPS spectra at varying temperatures, researchers can obtain valuable information about the kinetics and thermodynamics of tautomeric conversions. Both theoretical and experimental studies have indicated that, in the gas phase, these molecules predominantly exist in the keto form. Their structures are displayed in Fig. \ref{jiegou}. Among these molecules, only hypoxanthine can exist in two tautomeric forms. To take into account the experimental temperature, the TD framework was employed in the simulation process to generate an optimal spectral profile. Especially for hypoxanthine, the equilibrium distribution of two different tautomers is determined based on their relative Gibbs free energies. The probability ratio, also known as the Boltzmann population ratios (BPRs) or Boltzmann factor, between two tautomers $i$ and $j$ when they are in equilibrium is expressed as:
\begin{equation}
P_{i j}= \frac{P_i}{P_j}=e^{\frac{\Delta G_{j i}}{k_\text{B} T}},
\end{equation}
where $P_i$ and $P_j$ represent the probabilities of the two tautomers, $k_\text{B}$ is Boltzmann constant, and $\Delta G_{j i}$ ($\Delta G_{j i}=G_{j}-G_{i}$) is the Gibbs free energy difference between tautomers $i$ and $j$ at temperature $T$. By considering the temperature-dependent $\Delta G_{j i}$ and using the Boltzmann factor $P_{i j}$, researchers can determine the relative populations of the two tautomers at a given temperature.

Although experimental O1s XPS spectra are available for these molecules,\cite{plekan_theoretical_2008, feyer_tautomerism_2009, castrovilli_experimental_2018, plekan_x-ray_2012} their spectral resolution often falls short when it comes to clearly delineating vibrational structures, especially in the cases of uracil\cite{feyer_tautomerism_2009} and 5F-uracil.\cite{castrovilli_experimental_2018} Nonetheless, each molecule exhibits conspicuous peak asymmetry, providing clear evidence of the impact of vibronic coupling effects.  On the other hand, experimentalists have also conducted theoretical XPS studies on each system, but these studies have primarily focused on purely vertical excitations. In contrast, the objective of our study is to generate high-precision vibrationally resolved XPS spectra. These spectra will serve as a basis for conducting a comprehensive analysis that encompasses dominant vibronic transitions, active vibrational modes, contributions of 0-$n$ transitions, structural changes resulting from core ionization, and the effects of temperature. By achieving these research goals, we aim to derive general principles related to the vibronic coupling of carbonyl oxygen in these molecules. This endeavor will provide valuable insights into the intricate interplay between electronic and vibrational dynamics, contributing to the advancement of knowledge in the field of X-ray physics and spectroscopy.

%---------------------------------------------------------------------------------
 \section{Computational methods} \label{sec:method}
%---------------------------------------------------------------------------------
 \subsection{Franck-Condon simulations}
 
In this work, both TI\cite{ruhoff_algorithms_2000, wei_vibronic_2022, wei_vibronic_2023, cheng_vibrationally-resolved_2022, hua_theoretical_2020} and TD\cite{yan_eigenstate-free_1986} methods were used to model and analyze the vibronic profiles. Franck-Condon simulations were performed by including the Duschinsky rotation\cite{duschinsky_1937} effect. The harmonic oscillator (HO) approximation is always assumed. The relationship between the normal coordinates of the ground state ($\mathbf{q}^\prime$) and the core-ionized state ($\mathbf{q}$) of the molecule can be described by the Duschinsky transformation equation: $\mathbf{q}^\prime=\mathbf{J}\mathbf{q}+\mathbf{k}$. Here, $\mathbf{J}$ represents the Duschinsky rotation matrix, while $\mathbf{k}$ denotes the normal coordinate displacement. The displacement $\mathbf{k}$ is associated with the dimensionless Huang-Rhys factors (HRFs), which are widely used for analyzing the vibronic fine structures. Specially, The HRF for the $i$-th excited-state vibrational mode, denoted as $S_i$, is defined as  $S_i=\frac{1}{2\hbar}\omega_i k_i^2$. Here, $\omega_i$ is the corresponding frequency of the $i$-th mode. To further characterize the electron-phonon coupling strength for the particular vibrational mode $i$, the square root of $S_i$ can be taken, yielding the parameter $\lambda_i =\sqrt{S_{i}}$. 

\subsubsection{Time-independent method}

In the TI method, the FC amplitude $\langle 0|0\rangle$ is calculated based on matrices $\mathbf{J}$ and $\mathbf{k}$,\cite{hua_theoretical_2020} and the amplitude $\langle 0|n\rangle$ is recursively calculated starting from $\langle 0|0\rangle$.\cite{sharp_franckcondon_1964, ruhoff_recursion_1994, ruhoff_algorithms_2000} The Franck-Condon factor is derived by squaring the FC amplitude. Stick spectra are convoluted using a Lorentzian line shape given by the equation:
\begin{equation}
\Delta\left(E ; E_{0 n}, \gamma\right)=\frac{1}{\pi} \frac{\gamma}{\left(E-E_{0 n}\right)^{2}+\gamma^{2}}.
\end{equation}
The atom-specific vibrationally-resolved XPS cross section is calculated by:
\begin{equation}
\sigma_\text{XPS}(E)=\sum_{n}\langle 0 \mid n\rangle^{2} \Delta\left(E ; E_{0 n}, \gamma\right).
\label{ti-sigma}
\end{equation}
Here, $\gamma$ denotes the half-width-at-half-maximum (hwhm), $E$ refers to the binding energy, and $E_{0 n}$ denotes the 0-$n$ vibronic transition energy, which is given by
\begin{equation}
E_{0 n} = E_{00}+\sum_{i} n_{i} \hbar \omega_{i},
\end{equation}
where $n_i$ is the vibrational quantum number of the $i$-th normal mode of the final state. The 0-0 vibronic transition energy $E$$_{00}$ is represented by
\begin{equation}
E_{00}^\text{DR}   =  I^\text{ad} + \bigtriangleup {\varepsilon}_{0}.
\label{eq:E00:DR}
\end{equation}
Here, $I^\text{ad}$ and $\bigtriangleup {\varepsilon}_{0}$ stand for the adiabatic ionization energies and the zero-point vibrational energy (ZPE) difference between the two states, respectively. $I^\text{ad}$ is calculated according to the $\Delta$Kohn-Sham ($\Delta$KS) scheme\cite{triguero_separate_1999}, which can be expressed as:
\begin{equation}
I^\text{ad}  =  E_\text{FCH}|_\mathbf{min\,FCH} - E_\text{GS}|_\mathbf{min\, GS} + \delta. \label{eq:IP:ad}  
\end{equation}
Here, $E_{\rm{GS}}$ and $E_{\rm{FCH}}$ represent the total energies at the optimized geometries $\mathbf{min\ GS}$ and $\mathbf{min\ FCH}$, respectively.  A uniform shift of $\delta$ = 0.4 eV is applied to account for differential relativistic effects associated with removing electrons from the O1s core orbital.\cite{triguero_separate_1999}
Additionally, we also calculated the vertical ionization potentials ($I^\text{vert}$) using
\begin{equation}
I^\text{vert} = E_\text{FCH}|_\mathbf{min\, GS} -  E_\text{GS}|_\mathbf{min\, GS}+ \delta.
\label{eq:IP:vt}
\end{equation}
    
\subsubsection{Time-dependent method}

On the other hand, in the TD framework, the spectral profile can be obtained  as follows:
\begin{equation}
\sigma_\text{XPS}(E, T)= \operatorname{Re} \int_{0}^{\infty} d t \exp \left[\imath\left(E-E_{0 0}+\imath\gamma\right) t\right] \sigma(t, T).
\end{equation}
An advantage of the TD framework is that the finite temperature effect can be naturally included without additional computational cost. The auto-correlation function at finite temperature $T$, $\sigma(t, T)$, can be explicitly evaluated as:\cite{yan_eigenstate-free_1986}
\begin{equation}
\label{sigma/T}
\sigma(t, T) = \left(|\boldsymbol{f(t)} |^{-1/2}\right){\exp[\textbf{d}^\top \boldsymbol{f(t)} \textbf{d}]},
\end{equation}
where the superscript $\top$ stands for the transpose and
\begin{equation}
\boldsymbol{\psi(t)}  =\frac{1}{4}\left(\textbf{a}_{+} \textbf{R}^{\prime} \textbf{b}_{-}+\textbf{a}_{-} \textbf{R} \textbf{b}_{+}\right)\left(\textbf{a}_{+} \textbf{R} \textbf{b}_{-}+\textbf{a}_{-} \textbf{R}^{\prime} \textbf{b}_{+}\right), 
\end{equation}

\begin{equation}
\boldsymbol{f(t)}  = -\textbf{R}^{\prime} \textbf{b}_{-}\left(\textbf{a}_{+} \textbf{R}^{\prime} \textbf{b}_{-}+\textbf{a}_{-} \textbf{R} \textbf{b}_{+}\right)^{-1} \textbf{a}_{-}, 
\end{equation}
with 

\begin{align}
\textbf{a}_{ \pm}& =1 \pm \exp (\imath \boldsymbol{\omega} t), \\
\textbf{b}_{ \pm}& =(\bar{\textbf{n}}+1) \pm \bar{\textbf{n}} \exp \left(\imath \boldsymbol{\omega}^{\prime} t\right), \\
\bar{\textbf{n}}& =\left[\exp \left(\hbar \boldsymbol{\omega}^{\prime} / \mathbf{k} T\right)-1\right]^{-1}, \\
\textbf{R}^{\prime} & = \left(\textbf{R}^{-1}\right)^{\top} .
   \end{align}
Here, $\textbf{R}$ represents the transformation matrix for the dimensionless coordinates, and $\textbf{d}$ denotes the displacement between the dimensionless coordinates in the two electronic states. They are related to $\textbf{J}$ and $\textbf{k}$ with elements given respectively by\cite{yan_eigenstate-free_1986}
\begin{align}
R_{i j} & =\left(\omega_{i} / \omega_{j}^{\prime}\right)^{1 / 2} (\textbf{J}^{-1})_{i j}, \\
d_{i} &  =\left(\frac{\omega_{i}}{\hbar}\right)^{1 / 2}\left(-\mathbf{J}^{-1} \mathbf{k}\right)_{i}.
   \end{align}

\subsection{Simulation details}

All electronic structure calculations were performed at the DFT level using the Gamess-US software package\cite{schmidt_general_1993, gordon_advances_2005} with the B3LYP functional.\cite{becke_density-functional_1988, becke_new_1993, lee_development_1988} The basis set settings remained consistent with previous calculations. \cite{wei_vibronic_2022, hua_theoretical_2020, wei_vibronic_2023} For ground-state optimization and frequency calculation, the cc-pVTZ basis set \cite{dunning_gaussian_1989, kendall_electron_1992} was utilized. In all other calculations, including energy computations and determination of excited state structures and frequencies, the triple-$\zeta$ quality individual gauge for localized orbital (IGLO-III) basis set\cite{kutzelnigg_iglo-method:_1990} and corresponding model core potentials (MCPs)\cite{sakai_model_1997, noro_contracted_1997, bsjp} were employed specifically for the core-ionized oxygen (O$^*$) and the other (non-excited) oxygen. 

Spectral simulations were performed by using the modified\cite{hua_theoretical_2020} DanaVib package\cite{DynaVib}, which reads the DFT results obtained from Gamess-US.\cite{schmidt_general_1993, gordon_advances_2005} In practical applications, all spectral positions are calculated as relative energies with the 0-0 transition energy set to zero. The raw spectra are then calibrated by adding the absolute 0-0 transition energy. For single tautomeric forms of the molecule (xanthine, caffeine, thymine, 5F-uracil, and uracil), the final total spectrum is obtained by summing the specific spectra of each atom. Different hwhm values were set for each molecule to best match the experimental spectra: 0.035 eV for xanthine, 0.04 eV for caffeine and hypoxanthine, 0.05 eV for thymine and 5F-uracil, and 0.056 eV for uracil. Here all hwhm values account for the combined influence of all factors (lifetime, instrument resolution, nuclear motion, environmental influences, etc.) on the experimental linewidth, which are smaller than the experimental O1s core hole hwhm lifetime of 0.086 eV.\cite{nicolas_lifetime_2012} The simulated spectra were shifted by +0.87 eV for xanthine, +0.76 eV for caffeine, +0.62 eV for thymine, +0.80 eV for 5F-uracil, +0.30 eV for uracil,  and +0.60 eV for hypoxanthine to ensure better comparison with corresponding experimental data.

%---------------------------------------------------------------------------------
\section{\label{sec:level3} Results at zero temperature}
%---------------------------------------------------------------------------------

%+++++++++++++++++++++++++++++++
\subsection{Ionization potential and ZPE changes ($\Delta {\varepsilon}_{0}$)}
%++++++++++++++++++++++++++++++

\begin{figure*}
\includegraphics[width=0.6\textwidth]{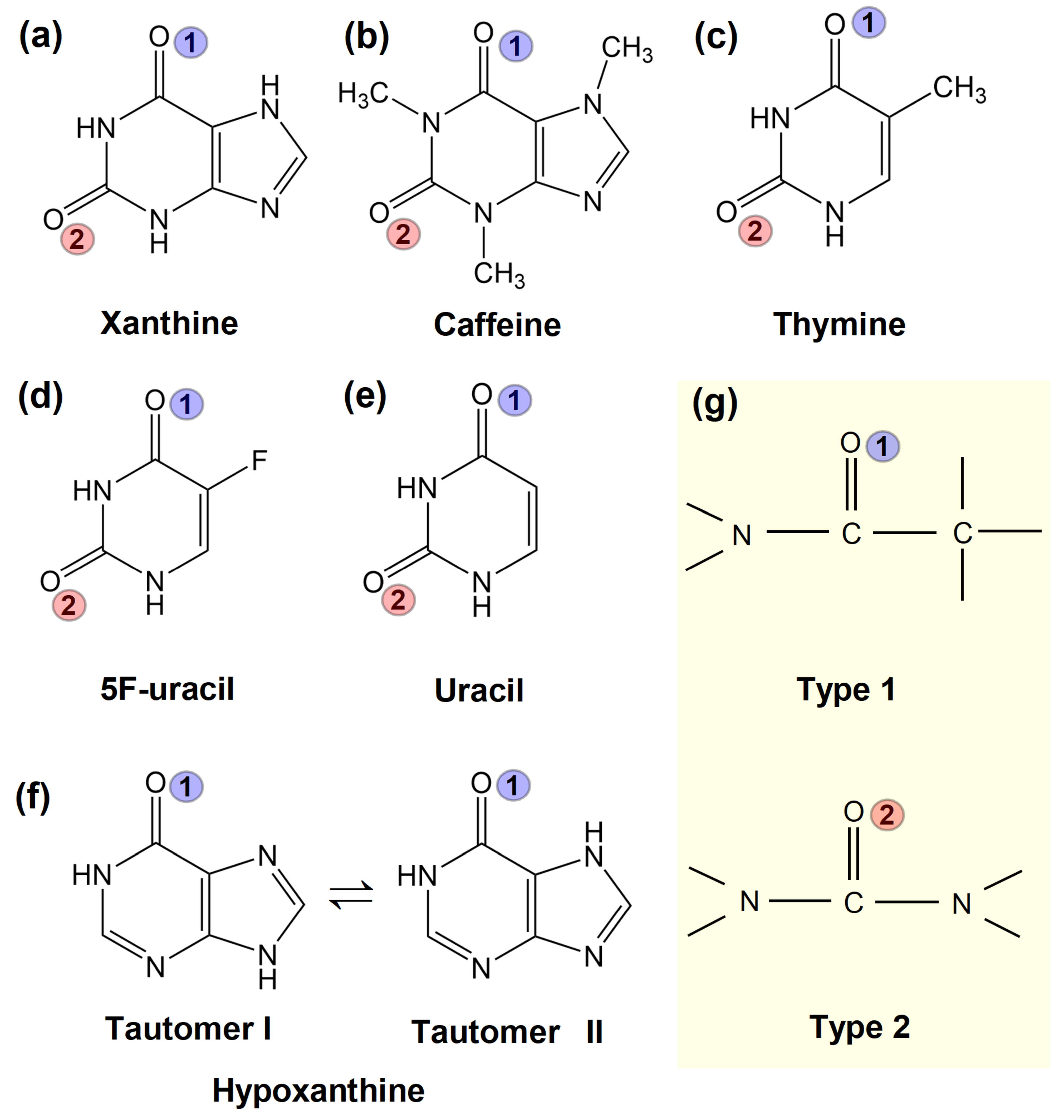}% Here is how to import EPS art
\caption{Structures of all systems under study: (a) xanthine, (b) caffeine, (c) thymine, (d) 5F-uracil, (e) uracil, and (f) two tautomers I and II of hypoxanthine. Each oxygen is labeled by 1 or 2 within a circle according to its local structure.  (g) Definitions for the two types of oxygens in panels (a)-(f).
}

\label{jiegou}
\end{figure*}

As shown in Table \ref{tab:ip}, the O1s ionization potentials (IPs) of all six molecules (depicted in Fig. \ref{jiegou}) are distributed within the range of 536.2--537.3 eV for vertical IPs and 535.7--536.8 eV for adiabatic IPs, spanning 1.1 eV in both cases. Our theoretical approach allows for a definitive differentiation, with variations ranging from 0.01 to 0.04 eV, between BEs of the two oxygen atoms in each system. These resolutions surpass what is typically achievable with experimental techniques. The analysis reveals that among the five single tautomeric molecules consisting of two oxygen atoms (xanthine, caffeine, thymine, 5F-uracil, and uracil), O2 in the {\chemNN} local structure consistently exhibits higher binding energies compared to O1 in the {\chemNC} local structure. This can be attributed to the higher electronegativity of N relative to C atoms. When compared with existing experimental data, our theoretical framework exhibits deviations ranging from 0.2 to 1.1 eV for vertical IPs or 0.4 to 1.5 eV for adiabatic IPs. Interestingly, these deviations align consistently with those observed in C/N 1s $\Delta$Kohn-Sham calculations.\cite{bagus_consequences_2016, pueyo_bellafont_validation_2015, pueyo_bellafont_prediction_2015, pueyo_SCF_performance_2016, du_theoretical_2022, wei_vibronic_2022, wei_vibronic_2023} 

BEs are highly sensitive to substituents, and our simulation accurately reproduces their relative values. The methyl group -CH$_3$ is a weak electron donating group (EDG).\cite{du_theoretical_2022}  We found that introducing -CH$_3$ significantly decreases the O1s BEs by comparing xanthine and caffeine [see structures in Figs. \ref{jiegou}(a) and (b). The O1 and O2 binding energies in xanthine were predicted to be 536.7 and 536.9 eV. The O1 and O2 binding energies of caffeine are 536.2 and 536.4 eV, respectively, reduced by -0.5 and -0.6 eV compared to xanthine. The two molecules show the same chemical shift of 0.2 eV between O1 and O2. Consistently, introducing the -CH$_3$ group can lead to a weak decrease in the O1s BEs by comparing uracil and thymine [see structures in Figs. \ref{jiegou}(e) and (c). BEs of O1 and O2 in uracil are 536.8 and 537.2 eV, respectively, with a chemical shift of 0.4 eV. While for thymine, the BEs are 536.7 and 537.0, respectively, and the chemical shift is 0.3 eV, with 0.1, 0.2, and 0.1 eV smaller than thymine. 

On the other hand, when the hydrogen atom is substituted with a fluorine atom, it leads to an increase of 0.4 eV in the O1 BE and 0.1 eV in the O2 BE of 5F-uracil as compared to uracil. The computed BEs for 5F-uracil are 537.1 eV for O1 and 537.3 eV for O2; while for uracil, the BEs are 537.3 eV for O1 and 537.2 eV for O2, respectively. These observations are plausible, and similar patterns have been found in N1s\cite{wei_vibronic_2023, du_theoretical_2022} or C1s\cite{mendolicchio_theory_2019} edge computations for other molecules.

In Table \ref{tab:ip}, the values of zero-point vibration energy difference ($\bigtriangleup {\varepsilon}_{0}$) between FCH and GS states are further predicted for all molecules. It is observed that these $\bigtriangleup {\varepsilon}_{0}$ values are small, falling within a range of 0.00--0.02 eV. This phenomenon has also been observed in azines,\cite{wei_vibronic_2022} which suggests that these molecules exhibit a weak Duschinsky rotation effect.

%+++++++++++++++++++++++++++++++
\subsection{Structural changes}
%++++++++++++++++++++++++++++++

\subsubsection{Global structural changes}
%++++++++++++++++++++++++++++++

The magnitude of the global structural alteration caused by O1s ionization can be supported by considering the total vibrational reorganization energy. According to the systems listed in Table \ref{si}, the range of reorganization energy resulting from the ionization of all core holes falls between 0.48 and 0.60 eV. In single tautomeric molecules, it is consistently observed that ionization on O2 leads to a greater reorganization energy (0.57--0.60 eV) compared to ionization on O1 (0.48--0.55 eV). This indicates that 1s ionization on O2 induces more significant structural deformation compared to ionization on O1.
    
%+++++++++++++++++++++++++++++++
\subsubsection{Local structural changes}\label{jiegoutaolun}
\begin{figure*}
\includegraphics[width=1.0\textwidth]{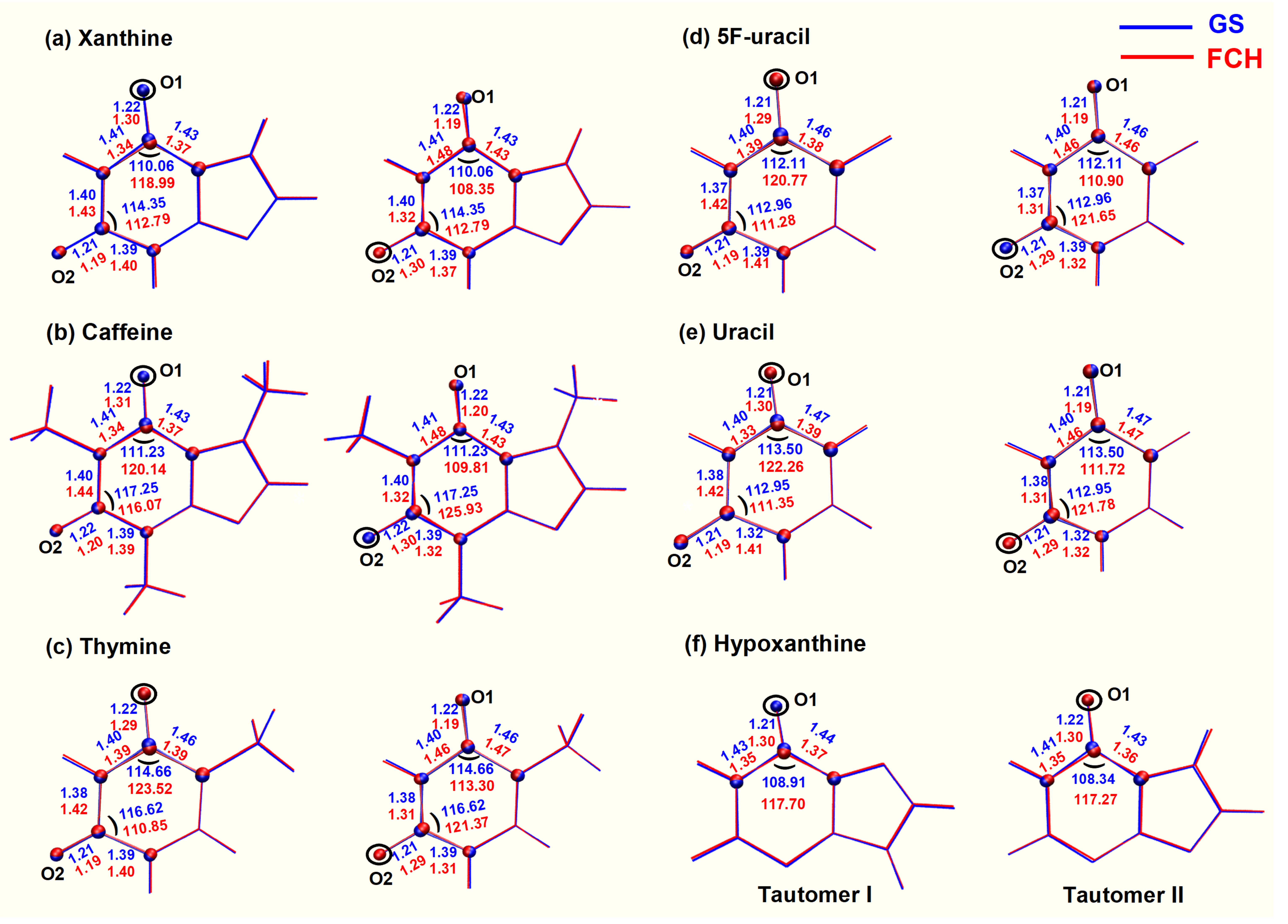}% Here is how to import EPS art
\caption{Comparison of geometries optimized in the GS and FCH states for all systems. Selected bond lengths (in {\AA}) and angles (in $^\circ$) of the carbonyl group involving O$^*$ and the remaining O are labeled. 
}\label{jiegou tol}
\end{figure*}

%++++++++++++++++++++++++++++++
Figure \ref{jiegou tol} provides a comparison of the optimized structures for all systems in both the ground and core-ionized states. This comparison specifically emphasizes the bond lengths and angles of the carbonyl groups where the core hole oxygen (O$^*$) and the remaining oxygen (O) are located. Table \ref{tab:str} offers a comprehensive summary of these structural alterations, revealing a consistent direction of change across each parameter. Interestingly, when molecules contain two oxygen atoms, O1s ionization generally induces opposite structural changes in carbonyl groups where O$^*$ and the remaining O are situated. For ease of expression, the carbon atom connected to O$^*$(O) is labeled C$^{\prime}$ (C$^{\prime\prime}$). The O$^*$-C$^{\prime}$ bond consistently experiences elongation by 0.08 to 0.09 {\AA}, while the O-C$^{\prime\prime}$ bonds uniformly contract by 0.02 to 0.03 {\AA}. For the carbonyl group containing O$^*$, the C$^{\prime}$-N/C$^{\prime}$-C bond distances consistently decrease by 0.01 to 0.08 {\AA}.  However, for the carbonyl group with the remaining O atoms, the C$^{\prime\prime}$-N bond lengths always increase, while the C$^{\prime\prime}$-C bond lengths remain mostly unchanged, with slight fluctuations ranging from -0.01 to +0.01 {\AA}. Regarding the bond angles, we observe an increase in all bond angles $\angle$N--C$^{\prime}$--$X$ ($X$=C or N) and a decrease in all bond angles $\angle$N--C$^{\prime\prime}$--$X$.  Furthermore, the alteration in $\angle$ N--C$^{\prime}$--$X$ (+4.8 to +8.9$^\circ$) is typically greater than in $\angle$ N--C$^{\prime\prime}$-$X$ ({-5.8} to {-1.2}$^\circ$). 

%+++++++++++++++++++++++++++++++
\subsection{Active vibrational modes}
%++++++++++++++++++++++++++++++
To elucidate the molecular structural changes induced by core ionization, active modes were analyzed based on the initial state. Vibrational modes in the ground state that exhibit large HRFs are referred to as active modes, and a threshold of $S_i$ $\geq$ 0.3 is applied for each system. Table \ref{si} illustrates the identification of 2--7 active modes responsible for the 1s ionization of O1 and O2 in these systems. A mode with large electron-phonon coupling  ($S_i$ $\geq$ 1) is identified as active in each atom-specific spectrum.

\begin{figure*}
\includegraphics[width=0.85\textwidth]{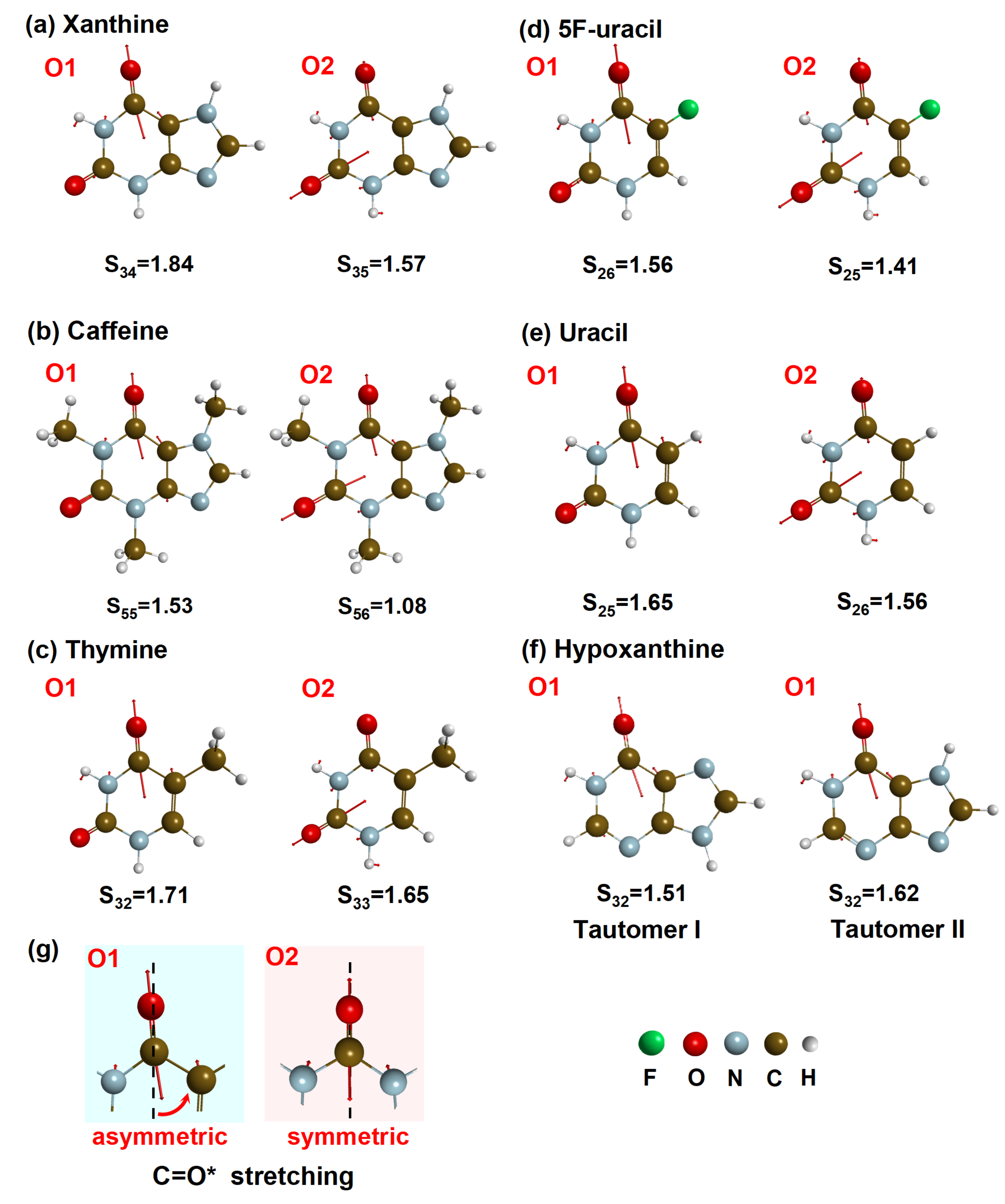}% Here is how to import EPS art
\caption{Visualization of the ground-state active vibrational modes for all systems with the largest HRF. Mode index and the corresponding HRFs are given.
}\label{mode tol}
\end{figure*}

Figure \ref{mode tol}(a)-(f) further depicts these strong electron-phonon coupling modes. It is discerned that these modes exhibit similar vibrational characteristics. The vibrations of each mode predominantly manifest through the stretching between O$^*$ and its adjacent C atoms, which strongly couples with the geometric reorganization induced by core ionization. As discussed in Section \ref{jiegoutaolun}, the ionization of O1s invariably results in the elongation of the O$^*$-C bond length.  All active modes are approximately assigned as the C=O$^*$ stretching vibrations. Those involving O1 and O2 show two categories of clear patterns with asymmetric and symmetric features, respectively, as summarized in Fig. \ref{mode tol}(g). The asymmetric pattern that deviates from the central C-O axis simply comes from different atoms (C and N) that are directly bonded with the carbonyl carbon.  Meanwhile, the structure basis for the symmetric pattern with respect to the C-O axis lies in the two nitrogens that are connected to the carbonyl carbon. 

%+++++++++++++++++++++++++++++++
\subsection{Atom- or tautomer-specific spectra at $T$=0 K}
%+++++++++++++++++++++++++++++++

Figure \ref{com_tol}(a)-(e) displays the simulated atom-specific XPS spectra of all five biomolecules with two oxygens using the TI method. In Fig. \ref{com_tol}(f), contributions from two tautomers of hypoxanthine are shown. The spectral profiles of the twelve O1s core holes exhibit remarkable consistency. The sum of FC factors rapidly converges to 0.99 at $n$=8 in most cases, except for caffeine, which only converges to 0.93 for O1 and 0.90 for O2 at $n$=6. Each system shows four distinct characteristic peaks for both O1 (or tautomer I) and O2 (or tautomer II), corresponding to the 0-0, 0-1, 0-2, and 0-3 transitions. Typically, the 0-2 transition has the highest intensity. This feature markedly contrasts with the previously examined C1s edges, where the spectra are often dominated by the 0-1 transition.\cite{hua_theoretical_2020} This difference is primarily due to the larger displacement of the PES caused by core ionization, as quantified by the vibrational recombination energy.\cite{qiu_narrowband_2021} Detailed assignments on stick vibronic transitions are provided in Fig. {\abctol}.  A threshold of FCFs ($F\geq 0.04$) was applied to filter out weak transitions, except for caffeine, whose threshold was set to 0.03. In each core hole, there are typically only 1--2 active vibronic transitions.

\section{Results at different temperatures} \label{sec:level4}
%---------------------------------------------------------------
%+++++++++++++++++++++++++++++++
\subsection{Total spectra for xanthine, caffeine, thymine, 5F-uracil, and uracil}
%+++++++++++++++++++++++++++++++

\subsubsection{Characteristics and interpretations of vibronic fine structures}

\begin{figure*}
\includegraphics[width=0.5 \textwidth]{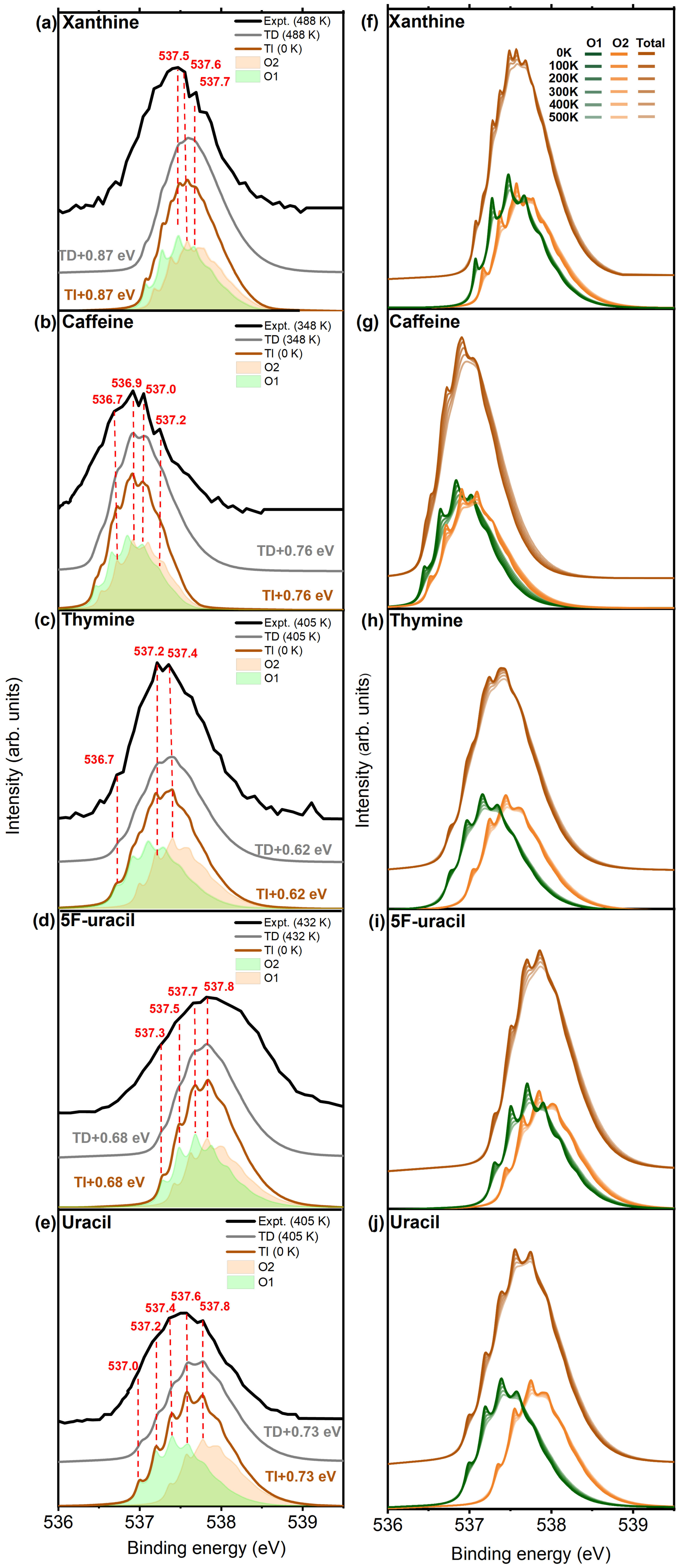}% Here is how to import EPS art
\caption{(a-e) Simulated vibrationally-resolved total O1s XPS spectra of xanthine, caffeine, thymine, 5F-uracil, and uracil using both the TD and TI methods. Theoretical spectra are uniformly shifted (indicated by numbers in eV) to 
better compare with the experiments (xanthine
\cite{plekan_x-ray_2012}, caffeine\cite{plekan_x-ray_2012},  thymine\cite{plekan_theoretical_2008}, 5F-uracil\cite{castrovilli_experimental_2018}, and uracil\cite{feyer_tautomerism_2009}). (f-j) The corresponding vibration-resolved O1s XPS spectra of each molecule in the temperature range of 0--500 K was simulated using the TD method.}
\label{xps_tol}
\end{figure*}

Figure \ref{xps_tol} presents the computed total O1s vibrationally-resolved XPS spectra of five single tautomeric form molecules: xanthine, caffeine, thymine, 5F-uracil, and uracil. In panel (a-e), the spectra were simulated both at the experimental temperatures (using the TD method) and at 0 K (using the TI method) for comparison purposes as well. Both theoretical spectral profiles are similar and show good agreement with experiments,\cite{plekan_x-ray_2012, feyer_tautomerism_2009, plekan_theoretical_2008, castrovilli_experimental_2018} significantly enhancing the accuracy of the spectra. Our computations reveal that the spectral curve at 0 K generally exhibits more pronounced peaks compared to those at experimental temperatures of several hundred K. This highlights the significant influence of temperature on the spectrum of these molecules.

\begin{figure*}
\includegraphics[width=1.0\textwidth]{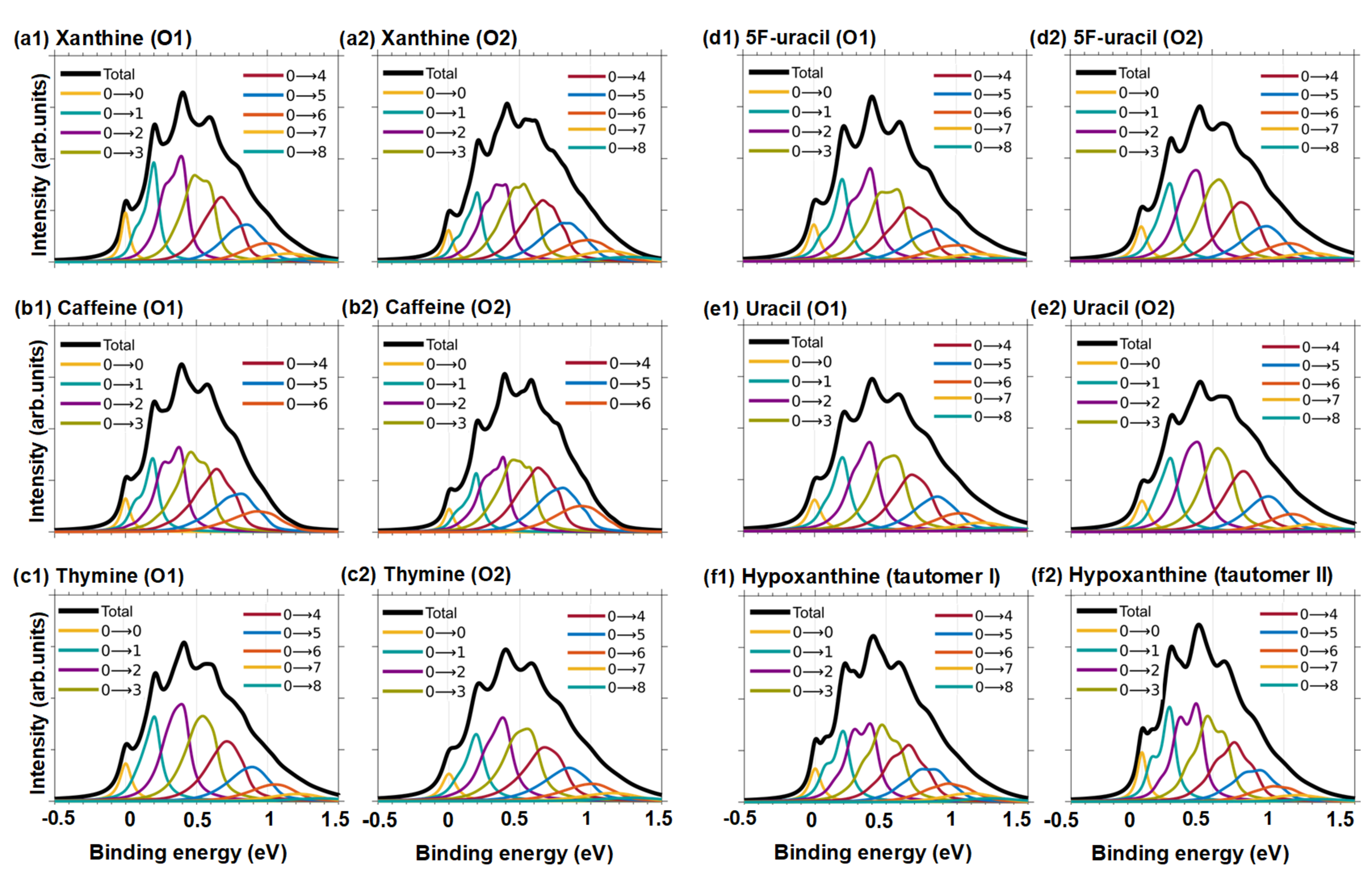}% Here is how to import EPS art
\caption{Contributions of different 0-$n$ transitions (until achieving convergence) for each oxygen, shown against the relative BE (0-0 transition energy set to zero): (a) xanthine, (b) caffeine, (c) thymine, (d) 5F-uracil, (e) uracil, and (f) hypoxanthine. Each spectrum was computed by the TI method at $T$=0 K. The same scale of the Y-axis is used.}
\label{com_tol}
\end{figure*}

As depicted in Fig. \ref{xps_tol}(a)-(e), the DR-TI method provides reliable assignments for these molecular spectra. In each molecule, O1 and O2 exhibit pronounced mixing between 536.0 and 539.0 eV, giving rise to rich spectral fingerprints.  The experimental spectrum of xanthine\cite{plekan_x-ray_2012} displays a prominent peak with considerable breadth at 537.5 eV, accompanied by a small feature at 537.7 eV.  As depicted in Fig. \ref{xps_tol}(a), our theoretical spectrum reproduces the overall profile and identifies three distinct peaks at 537.5, 537.6, and 537.7 eV, corresponding to the 0-2 (O1), 0-2 (O2), and 0-3 (O2) transitions, respectively [Fig. \ref{com_tol}(a)].  The experimental spectrum of caffeine\cite{plekan_x-ray_2012} exhibits four fingerprints at 537.6, 536.9, 537.0, and 537.2 eV. Our simulations capture the fine structure in Fig. \ref{xps_tol}(b) and attribute them to the 0-1 (O2), 0-2 (O2), 0-3 (O1), and 0-4 (O1) transitions [see Fig. \ref{com_tol}(b)].

As shown in Fig. \ref{xps_tol}(c), the experimental spectrum of thymine\cite{plekan_theoretical_2008} exhibits a pair of minor peaks at 537.2 and 537.4 eV, with a separation of 0.2 eV. Our theoretical simulation perfectly matches the experimental results, and both identified features are attributed to O2 contributions arising from the 0-1 and 0-2 transitions, respectively.  Additionally, a small peak is observed in the lower energy spectral region at 536.7 eV, corresponding to the 0-0 transition of O1 [see Fig. \ref{com_tol} (c)].

Figures \ref{xps_tol}(d) and (e) present the simulated O1s XPS spectra for 5F-uracil and uracil. The experimental spectra of 5F-uracil\cite{castrovilli_experimental_2018} and uracil\cite{feyer_tautomerism_2009} exhibit asymmetric broad peaks at 537.8 and 537.6 eV respectively, with no obvious vibronic features. Theoretical calculations reveal four distinct characteristics for 5F-uracil, with three peaks primarily originating from O1 (537.3, 537.5, and 537.7 eV, corresponding to the 0-0, 0-1, 0-2 transitions, respectively), and the fourth peak attributed to O$_2$ (537.8 eV for the 0-2 transitions) [Fig. \ref{com_tol} (d)]. For uracil, we predict five prominent fingerprints distributed within the range of 537.0--537.8 eV, with adjacent intervals of 0.2 eV. Peaks 1--4 are attributed to O1 (0-0, 0-1, 0-2, and 0-3 transitions), while peak 5 is caused by O2 (0-2 transitions) [see Fig. \ref{com_tol} (e)].

\subsubsection{Influence of temperature}

Figure \ref{xps_tol} (f)-(j) illustrates the spectral profiles of the five single tautomeric form systems at different temperatures. Temperature variations consistently affect the spectral characteristics of these molecules. As the temperature increases, the intensity of each fine feature within the atom-specific spectrum of each molecule progressively diminishes. Given that each total spectrum is composed of the sum of two atom-specific spectra, the intensity of each vibronic feature contributing to this total spectrum correspondingly diminishes.

%+++++++++++++++++++++++++++++++
\subsection{Total spectra for hypoxanthine}
%+++++++++++++++++++++++++++++++

\subsubsection{Spectra at the experimental temperature}

\begin{figure*}
\includegraphics[width=1.0\textwidth]{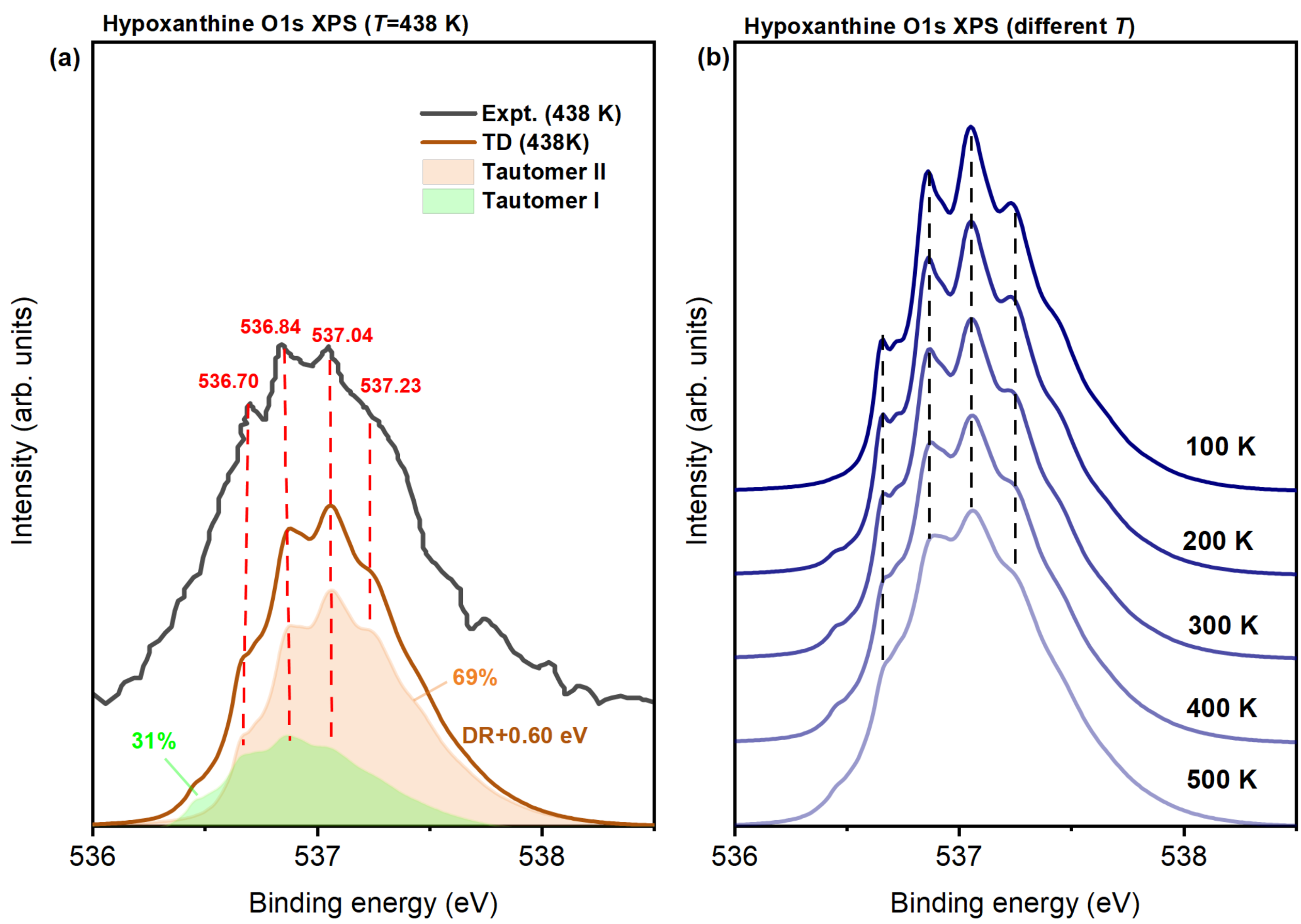}% Here is how to import EPS art
\caption{Vibrationally-resolved O1s XPS spectra of hypoxanthine simulated using the TD method at (a) 438 K (experimental\cite{plekan_x-ray_2012} temperature) and (b) 100--500 K.  To better compare with the experiment, all theoretical spectra are uniformly shifted by +0.60 eV.
}\label{hy_tem}
\end{figure*}

To account for the proportions of tautomers at a specific temperature, only the TD method is employed for the simulation of the total spectrum of hypoxanthine. Thermochemical calculations are performed to determine the difference in Gibbs free energy and calculate the BPRs for tautomer II and tautomer I forms\cite{farrokhpour_theoretical_2011}, which are found to be 0.6895 and 0.3105, respectively, at the experimental spectral temperature of 438 K.\cite{plekan_x-ray_2012}   Figure \ref{hy_tem}(a) illustrates the corresponding total O1s XPS spectra of hypoxanthine,\cite{plekan_x-ray_2012} revealing a strong mixed spectrum of tautomers I and II. Our simulation agrees well with the experimental results and accurately reproduces all four characteristic peaks at 536.7, 536.8, 537.0, and 537.2 eV.

\subsubsection{Influence of temperature}
Table \ref{bpr} presents the proportions of the two tautomers of hypoxanthine at different temperatures. As the temperature increases, the proportion of tautomer I gradually increases, and this should be taken into account during spectral assignment (see Fig. {\hytem}). Figure \ref{hy_tem}(b) further illustrates the total vibration-resolved XPS spectra of hypoxanthine at different temperatures using the TD method. The spectral profile is influenced by the escalating temperature, which predominantly gives rise to a gradual diminution in the intensity of the characteristic peaks.

\section{Summary and Conclusions}
%---------------------------------------------------------------------------------
In summary, we have calculated vibrationally-resolved O1s XPS spectra for six small bio-molecules (xanthine, caffeine, thymine, 5F-uracil, uracil, and hypoxanthine) employing both the TI and TD methods. Our analysis shows that all binding energy and high-resolution spectra achieved good agreement with existing experimental data, thereby confirming the reliability of our analysis. We have further analyzed the binding energy, geometric changes induced by core ionization, active vibrational modes, and spectral features for each molecule. Specifically, the TI method has been utilized to elucidate the vibronic transitions in the spectra. In contrast, the TD method is designed to provide accurate spectra at the experimental temperature by taking into account the Boltzmann population ratios of the tautomers. Our study provides a comprehensive understanding of the vibronic coupling properties exhibited by these molecules, offering valuable insights into their intricate dynamics and interactions.

For the six cyclic molecules studied, specific rules on vibronic properties have been derived. Notably, we have observed that the presence of a C-N group in the carbonyl group enhances the O1s BE and the total vibrational reorganization energy. Additionally, our predictions indicate small $\bigtriangleup {\varepsilon}_{0}$ values for each core hole, indicating a weak Duschinsky rotation effect in these molecules. For molecules with single tautomeric forms, we found that 1s ionization in O2 atoms results in more significant global structural changes, as expressed by the total vibrational reorganization energies, as compared to O1. O1s ionization always leads to an increase in the bond angle $\angle$ N-C$^{\prime}$-$X$ ($X$=C, N) and the bond length O$^*$-C$^{\prime}$. Furthermore, we have identified an active vibrational mode involving O$^*$-C stretching, which is strongly coupled to these structural changes. It is worth noting that the spectral curves of these molecules are temperature-sensitive. Specifically, an increase in temperature leads to a decrease in the intensity of characteristic peaks in each total vibrationally resolved spectrum.

%--------------------------------------------------------------------------------
\section{Acknowledgments}
%--------------------------------------------------------------------------------
Financial support from the National Natural Science Foundation of China (Grant No. 12274229)  is greatly acknowledged. M.W. thanks to Fund for Fostering Talented Doctoral Students of Nanjing University of Science and Technology.

%---------------------------------------------------------------------------------
%\bibliography{O1s}% Produces the bibliography via BibTeX.

\clearpage
 
%========================TABLES %========================

% ----  [TABLE 1: tab:fun] ----
\begin{table*}[]
    \centering
        \caption{
Vertical and adiabatic ionization potentials ($I^\text{vert}$ and $I^\text{ad}$), 0-0 transition energies ($E^\text{DR}_\text{00}$), and $\bigtriangleup {\varepsilon}_{0}$ for all six molecules simulated using B3LYP method.  The calculated values are compared with experimental data, and relative deviations are provided in parentheses. All energies are in eV.
} \label{tab:ip}

    \resizebox{\textwidth}{!}{
\begin{ruledtabular}
\begin{threeparttable}
    \begin{tabular}{lcccccc}

         Molecule & Core & Expt. & $I^\text{vert}$ & $I^\text{ad}$ & $E^\text{DR}_\text{00}$ & $\bigtriangleup {\varepsilon}_{0}$  \\ \hline
       % H$_2$O & ~ & 539.79\tnote{a} & 539.61 (-0.18) & 539.44 (-0.35) & 539.50 & -0.06  \\ 
        Xanthine & O1& 537.45\tnote{a} & 536.74 (-0.71) & 536.20 (-1.25) & 536.20 & 0.00  \\ 
        &O2 & 537.45\tnote{a} &536.87 (-0.58) & 536.30 (-1.15) & 536.29 & 0.01   \\ 
        Caffeine & O1 & 536.85\tnote{a} & 536.23 (-0.62) & 535.69 (-1.16) & 535.68 & 0.01 \\ 
        &O2 & 536.85\tnote{a} & 536.35 (-0.50) & 535.76 (-1.09) & 535.75 & 0.01 \\ 
        Thymine & O1 & 537.30\tnote{b} & 536.67 (-0.63) & 536.14 (-1.16) & 536.13 & 0.01  \\ 
        &O2 & 537.30\tnote{b} & 536.96 (-0.34) & 536.41 (-0.89) & 536.41 & 0.00  \\ 
        5F-uracil & O1 &  537.56\tnote{c}  & 537.14 (-0.42) & 536.61 (-0.95) & 536.60 & 0.01  \\ 
        &O2 & 538.22\tnote{c}  & 537.30 (-0.92) & 536.76 (-1.46) & 536.75 & 0.01  \\ 
        Uracil & O1 & 537.60\tnote{d} & 536.79 (-0.81) & 536.25 (-1.35) & 536.24 & 0.02  \\ 
       & O2 & 537.60\tnote{d} &537.17 (-1.05) & 536.61 (-0.99) & 536.61 & 0.00  \\ 
     Hypoxanthine (I) & O1& 537.00\tnote{a} & 536.34 (-0.66) & 535.84 (-1.16) & 535.82 & 0.02 \\ 
     Hypoxanthine (II) & O1& 537.00\tnote{a} & 536.54 (-0.46) & 536.05 (-0.95) & 536.04 & 0.01   \\ 
    \end{tabular}
\begin{tablenotes}
%\item[a] Sankari et al.\cite{sankari_vibrationally_2003}
\item[a] Plekan et al.\cite{plekan_x-ray_2012}
\item[b] Plekan et al.\cite{plekan_theoretical_2008}
\item[c] Castrovilli et al.\cite{castrovilli_experimental_2018}
\item[d] Feyer et al. \cite{feyer_tautomerism_2009}
\end{tablenotes}

\end{threeparttable}
\end{ruledtabular}
    }

\end{table*}

%%%%%%%%%%%%%%%%%%%%%%%Table Er
\begin{table}
    \centering
\caption{
Analysis of selected vibrational modes with large Huang-Rhys factors (with a threshold $S_i\geq 0.3$) for all six molecules. The vibrational frequencies $\omega_i$  and the total vibrational reorganization energy $E_r$ of the excited state (FCH) are given.}   
    \begin{ruledtabular}
\begin{threeparttable}
    \begin{tabular}{lccccc}
Molecule &O$^*$& $i$& $\omega_i$ (cm$^{-1}$)& $S_i$& $E_r$ (eV)   \\     \hline
Xanthine   &   O1 & 11 & 627.6 & 0.47& 0.53   \\  
           &      & 34 & 1732.5 & 1.84&       \\
           &   O2& 35 & 1862.1 & 1.57& 0.60   \\ \hline  
Caffeine   &   O1 &20  & 643.7 & 0.50 & 0.55 \\
           &      & 55 & 1709.9 &1.53 & \\
           &      &56  & 1815.2 & 0.41& \\ 
           &   O2 & 20 & 633.4  &  0.31& 0.60\\
           &      & 29 & 1007.2 & 0.43 &    \\
           &      & 55 & 1678.4 & 1.08 &   \\
           &      & 56 & 1827.0 & 0.78 &  \\ \hline   
Thymine    & O1 & 32  & 1666.1&1.71  & 0.51 \\  
           &   O2 &33   &1846.6 & 1.65 &0.57  \\ \hline
5F-uracil  & O1&26   & 1693.7 & 1.56 &0.49 \\ 
           & O2  & 26  & 1719.4& 0.38 & 0.57 \\
           &      &27   & 1869.2& 1.41 &     \\ \hline 
Uracil     & O1 & 25  & 1651.2 &1.65 & 0.48 \\
           & O2  & 14  & 917.2 & 0.35 & 0.59 \\ 
           &      & 18  & 1164.7& 0.30 &    \\
           &       &26   &1870.0 & 1.56 &   \\ \hline   
    Hypoxanthine (I)& O1  & 5 & 524.1  & 0.33 & 0.48 \\
               &    &12 &  730.1 & 0.39 &    \\ 
               &    &32 & 1709.7 & 1.51 &    \\ 
Hypoxanthine (II)& O1  & 5 &  524.5  &  0.35 & 0.49 \\
                &   &32 & 1740.1 &  1.62  &    \\  
    \end{tabular}
%    \begin{tablenotes}
%         \end{tablenotes} 
         \end{threeparttable}
\end{ruledtabular}
    \label{si}
    \end{table}

%\weijie{请确认非激发的氧原子几个，改一下单复数} \wmr{就一个未激发的氧}
% ----  [TABLE 2: tab:IP] ----
\begin{table*}[]
    \centering
   \caption{Structural changes of all molecules at their optimized excited state geometry (\textbf{min FCH}) as compared with the optimized ground state geometry  (\textbf{min GS}). Selected bond lengths (in {\AA}) and angles (in $^\circ$) near the ionized (O$^*$) and the other (non-excited) oxygen are presented in Fig. \ref{jiegou tol}. The carbon atom bonded with O$^*$ is denoted as C$^{'}$, while C$^{''}$ signifies the carbon atom bonded to the other oxygen.}    
    \resizebox{\textwidth}{!}{
\begin{ruledtabular}
\begin{threeparttable}
    \begin{tabular}{lccccccccc}
         Molecule & Core & O$^*$--C$^{'}$ & C$^{'}$--N & C$^{'}$--C & O--C$^{''}$ & C$^{''}$--N &  C$^{''}$--C & $\angle $N--C$^{'}$--X & $\angle$ N--C$^{''}$--X \\ \hline
 Xanthine & O1 & +0.09 & -0.07 & -0.07 & -0.02 & +0.03, +0.01 & & +8.93 & -1.56  \\ 
 &O2& +0.08 & -0.08, -0.07 &  & -0.03 & +0.07 & -0.00 & +8.40 & -1.71   \\ 
 Caffeine & O1 & +0.09 & -0.08 & -0.06 & -0.02 & +0.03, +0.01 &  & +8.91 & -1.18  \\ 
 &O2& +0.09 & -0.08, -0.07 &  & -0.02 & +0.07 & -0.00 & +8.68 & -1.42  \\ 
 Thymine & O1 & +0.08 & -0.01 & -0.07 & -0.02 & +0.03, +0.02 &  & +8.86 & -5.77  \\ 
 &O2& +0.08 & -0.07, -0.07 &  & -0.03 & +0.07 & +0.01 & +4.75 & -1.36  \\ 
 5F-uracil & O1 & +0.08 & -0.01 & -0.07 & -0.02 & +0.03, +0.02 &  & +8.72 & -1.68  \\ 
 &O2 & +0.08 & -0.07, -0.07 &  & -0.03 & +0.07 & +0.00 & +8.69 & -1.21  \\ 
 Uracil & O1 & +0.08 & -0.07 & -0.08 & -0.03 & +0.04, +0.02 &  & +8.76 & -1.60 \\ 
 &O2 & +0.08 & -0.07, -0.07 &  & -0.03 & +0.06 & +0.01 & +8.76 & -1.78   \\ 
  Hypoxanthine (I) & O1 & +0.08 & -0.08 & -0.07 & ~ & ~ & ~ & +8.79 &   \\ 
 Hypoxanthine (II) & O1 & +0.08 & -0.06 & -0.07 & ~ & ~ & ~ & +8.93&   \\ 
    \end{tabular}
    \end{threeparttable}
\end{ruledtabular}
    \label{tab:str}
    }
\end{table*}

%========================TABLES %========================
%[\weijie{See Eq. (ref XXX)} ]
% ----  [TABLE 1: tab:fun] ----
\begin{table}[]
    \centering
        \caption{Simulated relative Gibbs free energies $\bigtriangleup$$G$ and the Boltzmann population ratios of tautomers I and II for hypoxanthine at different temperatures.} 
        \label{bpr}

\begin{ruledtabular}
\begin{threeparttable}
    \begin{tabular}{ccccc}
                     &                   &\multicolumn{2}{c}{BPR(\%)} &  \\
                                 \cmidrule{3-4}
        Temperature (K) & $\bigtriangleup$$G$ (eV) & Tautomer I & Tautomer II  \\ \hline
         % 0 K & 2.9148  & 0.00 & 100.00 \\
        100  & 2.9088  & 2.93 & 97.07  \\ 
        200  & 2.9128  & 14.78 & 85.22 \\ 
        300  & 2.9158  & 23.70 & 76.30  \\ 
        400  & 2.9098  & 29.42 & 70.58 \\ 
        438  & 2.9048  & 31.05 & 68.95 \\
        500  & 2.8938  & 33.26 & 66.74 \\ 
    \end{tabular}

\end{threeparttable}
\end{ruledtabular}
\end{table}

\clearpage

%----------------- Fig. 1(结构图)

%----------------- Fig. 2(jiegou tol)
%----------------- Fig. 3(mode tol)

%----------------- Fig. 4(com_tol)
% within each system\weijie{看不懂后半句}\wmr{每个系统内中非等价氧原子不同0-$n$跃迁直至实现收敛的贡献分析}
%\weijie{确保你的不同图的y轴的范围是一致的}\wmr{已改}
%\weijie{1、图字体实在太小，特别是X轴 2、建议标号改成a1，a2，b1，b2，c1，c2，d1，d2} \wmr{已改}

%----------------- Fig. 5( xps_tol)

%=========================FIGURES===================================
%----------------- Fig. 6(hy--tem)

%\nocite{*}
%\end{CJK*}
\end{document}